\documentclass[acmsmall]{acmart}

\usepackage{multirow}
\usepackage{subcaption}
\usepackage{color}

\usepackage{verbatim}
\newcommand{\newt}[1]{{#1}}

\begin{document}

\setcopyright{acmlicensed}
\acmJournal{PACMHCI}
\acmYear{2022} \acmVolume{6} \acmNumber{CSCW2} \acmArticle{541} \acmMonth{11} \acmPrice{15.00}\acmDOI{10.1145/3555599}

\title[The Role and Dynamics of Visual Content in the 2020 U.S. Election Misinformation Campaign]{%
Stop the [Image] Steal: The Role and Dynamics of Visual Content in the 2020 U.S. Election Misinformation Campaign
}

\author{Hana Matatov}
\email{hanama888@gmail.com}
\orcid{0000-0001-5826-8921}
\affiliation{
  \institution{Technion - Israel Institute of Technology}
  \city{Haifa}
  \country{Israel}}
\author{Mor Naaman}
\email{mor.naaman@cornell.edu}
\orcid{0000-0002-6436-3877}
\affiliation{
  \institution{Cornell Tech}
  \city{New York}
  \state{New York}
  \country{United States}}
\author{Ofra Amir}
\email{oamir@technion.ac.il}
\orcid{0000-0003-2303-3684}
\affiliation{
  \institution{Technion - Israel Institute of Technology}
  \city{Haifa}
  \country{Israel}}

\begin{abstract}
   Images are powerful. Visual information can attract attention, improve persuasion, trigger stronger emotions, and is easy to share and spread. We examine the \newt{characteristics of the popular images shared} on Twitter as part of ``Stop the Steal'', the widespread misinformation campaign during the 2020 U.S. election. We analyze the spread of the forty most popular images shared on Twitter as part of this campaign. Using a coding process, we categorize and label the images according to their type, content, origin, and role, and perform a mixed-method analysis of these images' spread on Twitter. Our results show that popular images include both photographs and text rendered as image. Only very few of these popular images included alleged photographic evidence of fraud; and none of the popular photographs had been manipulated. Most images reached a significant portion of their total spread within several hours from their first appearance, and both popular- and less-popular accounts were involved in various stages of their spread.   
\end{abstract}




\maketitle
\section{Introduction}
\label{introduction}

Visual content plays a significant role in online misinformation~\cite{firstdraftInformationDisorder,highfield2016instagrammatics,marwick2017media,paris2019deepfakes,OutofContextPost}. 
This work presents a case study of the ``Stop the Steal'' misinformation campaign before, during, and following the 2020 U.S. elections to expand our understanding \newt{of the nature and dynamics of popular images shared in such a campaign.} 
This campaign, unfolding over social media platforms including Facebook and Twitter, spread unsubstantiated and misleading claims of voter fraud, allegations that the election was stolen, and calls to overturn the results~\cite{nytimestweets,FindVotesGeorgia,callElectionCorrupt, benkler2020mail, CenterforanInformedPublic2021}.
These widespread claims influenced public opinion~\cite{Berlinski} to such an extent that, as of December 2020, approximately a third of the US population did not trust the election results~\cite{trustpoll, benkler2020mail}, a number that remained similar nearly a year after the election~\cite{AmericanValuesSurvey}.

We focus here on visual content, given the persuasive nature of images~\cite{zannettou2020characterizing,wittenberg2021minimal,paris2019deepfakes,hsieh2011different,counts2011taking,highfield2016instagrammatics} and the frequent use of visual information in such campaigns~\cite{types_election_misinformation,marchal2021investigating,corrigall2012power,casas2019images,bhargava2020mapping,bebic2018not,zakem2018exploring,ling2021dissecting}.
Because of the persuasive power of images and memes~\cite{wittenberg2021minimal,paris2019deepfakes,ling2021dissecting,hsieh2011different,graber1996say,counts2011taking,casas2019images}, visual content may be used to convey powerful messages and provide more ``believable'' evidence or indication that certain activities occurred.
Images also provide a research opportunity because, similar to URLs, they provide a method for matching stories about the same subject or topic~\cite{matatov2022dataset,dubey2018memesequencer,moreira2018image,zannettou2018origins}.
For example, when using a Twitter dataset, images give us a robust mechanism to associate all tweets that shared the same or a very similar image, regardless of text similarity.

We use data from the Stop the Steal campaign on Twitter, and look at the forty most popular images shared during that campaign. The source of our data is the VoterFraud2020 dataset~\cite{abilov2021voterfraud2020}, which includes over 33 million tweets and retweets containing key phrases related to voter fraud claims, surrounding the U.S. 2020 election, between October 23rd and December 16th, 2020.
The dataset includes 167,696 perceptual hash (pHash) values~\cite{monga2006perceptual} for images appearing in the tweets. The pHash values allow matching of identical images across tweets. 
The forty most popular images were all shared in the dataset more than 1,000 times each, with the top two images appearing in the data in over 10,000 tweets and retweets (each).

Our goal is to \newt{examine the spread of images as part of the Stop the Steal campaign on Twitter} by using an analysis of the top images shared and \newt{characteristics of} their spread. 
We set out to characterize (1) the types and roles of the top images that were shared in this campaign; (2) the temporal patterns of the images' spread; and (3) the role of different users with different network prominence in sharing and popularizing the images.

To this end, we take
a mixed-methods approach to analyzing the popular images. Our analysis uses a qualitative coding process to categorize the images along multiple categories, including  type, origin, content, and role.
We further conduct a quantitative data-driven analysis of how the images were shared in the campaign.

Our results expand the understanding of how images are used in large-scale misinformation campaigns, and highlight the challenges these patterns pose in addressing the misinformation problem. 
We show that only a small number of the popular images attempted to provide ``evidence'' for voter fraud, and that none of the popular images were photographs that were manipulated in any way. In fact, many of the popular ``images'' shared were simply text which was rendered as an image to highlight its contents, and only one was a ``meme'' (a captioned picture that tends to spread widely online)~\cite{miltner2018internet,gal2016gets,ling2021dissecting}.
Most images in this set of top~40 had similar temporal distribution patterns, regardless of their labels.
The images spread quickly: almost 50\% of the images reached the half-life of their spread (time until half of their shares in our data) within only three hours from when they were first found in our data.
\newt{We show that the spread of these images had a mixed-participation pattern with accounts of varying popularity participating during all stages of the images' propagation. }
\newt{This result is consistent with the findings of earlier works showing the multi-directional nature of participation in the campaign between accounts with different levels of popularity~\cite{CenterforanInformedPublic2021}}. 
At the same time, our results show that images that offered ``evidence'' of voter fraud, i.e. more directly controversial
claims, were less readily shared by larger accounts (1,000-100,000 followers) compared to smaller ones.
Finally, we present a temporal order-based analysis that shows that amongst popular users, some repeat participants were particularly active in the early spread of multiple instances of images in this campaign -- again supporting earlier findings~\cite{CenterforanInformedPublic2021, gallagher2021sustained}.

\newt{The results provide new insights about the use of visual content during misinformation campaigns, and offer implications for designing solutions to cope with such content, e.g., moderation of content or accounts~\cite{gillespie2018custodians, papakyriakopoulos2020spread,Conger2021}.
For example, the identification of the most prevalent images and their characteristics shows that the most prevalent issue is, at least at this point, \textit{not} image manipulation, though this topic received significant policy and research attention~\cite{paris2019deepfakes,miltner2018internet,zakem2018exploring,chesney2019deep,gal2016gets,ling2021dissecting,marra2018detection,marra2019gans}.
Our findings emphasize the importance of addressing visual misinformation in the form of ``out-of-context'' images, where images are shared in a context different from that in which they were created~\cite{OutofContextPost,matatov2022dataset,firstdraftInformationDisorder,matatovdejavu,moreira2018image,gupta2013faking}.
Further, 40\% of the popular images were actually text or quotes rendered as images, suggesting that extracting and analyzing image text could also be an important tool for addressing misinformation.
The quick spread of the popular images demonstrates that any effective platform approach to mitigating the spread of misinformation would need to be applied within a few hours from the time the image was first shared.
Finally, the mixed-participation pattern shows the challenges of using selective user suspension as a way to address misinformation. However, we show that repeat and early offenders exist within the more popular users, which may suggest a strategy for selective yet effective enforcement.}

\section{Background}
\label{background}

In this work, we study the role and dynamics of visual content in a misinformation campaign using the Stop the Steal misinformation campaign surrounding the 2020 U.S. election as a case study.
We therefore build on two general bodies of work: related work in the field of visual communication, and in particular visual misinformation (Section~\ref{subsec:visual_background}); and  prior work examining misinformation in election campaigns, in particular in the context of the recent U.S. elections (Section~\ref{subsec:election_background}).%

\newt{Recent social computing research efforts have focused on examining misinformation and disinformation campaigns online~\cite{lazer2018science,xiao2021sensemaking,vosoughi2018spread}, and were especially interested in understanding the loosely-coordinated nature of such campaigns~\cite{arif2018acting,jakesch2021trend,stewart2018examining,starbird2014rumors,starbird2019disinformation}. 
In particular, Starbird et al.~\cite{starbird2019disinformation}
call for considering ``strategic information	operations'' as a critical concern for CSCW researchers, highlighting the organic aspects of such campaigns, often reflecting the activities of diverse actors who are implicitly coordinated. Similar dynamics had been observed in the Stop the Steal campaign that we study in this work~\cite{abilov2021voterfraud2020,CenterforanInformedPublic2021}.
Misinformation also received attention from social computing researchers looking at interventions that can help people avoid or detect it~\cite{spezzano2021s,micallef2021fakey,zhang2018structured,bhuiyan2021nudgecred,jahanbakhsh2021exploring}.
For example, in~\cite{jahanbakhsh2021exploring}, the authors investigated the design of lightweight interventions %
that nudge users to assess the accuracy of information before sharing it.
Other ways social computing research had considered addressing misinformation included moderation~\cite{seering2020reconsidering,gillespie2020content,jhaver2018online}.
These papers acknowledge the problems of online harassment and misinformation spread, discuss the extent of the problem, and suggest potential moderation tools -- not directly studied in our work here, but related to its implications.}

\subsection{Visual Content and Misinformation}
\label{subsec:visual_background}

Visual content can help attract attention~\cite{hsieh2011different,counts2011taking}; be more persuasive in providing evidence that something happened~\cite{zannettou2020characterizing,wittenberg2021minimal,paris2019deepfakes}; trigger  stronger emotional reactions~\cite{graber1996say}; and increase or speed up the propagating of content due to images’ ease of share~\cite{highfield2016instagrammatics,zannettou2018origins,zannettou2020characterizing}.
Research had in particular examined these advantages in relation to other formats, like text~\cite{hsieh2011different,wittenberg2021minimal}.
For example, research had shown that attention to web content with visual imagery is much higher compared to text-only~\cite{hsieh2011different};
and that visual content (in this case, video) increases the believability of the presented content compared to textual form~\cite{wittenberg2021minimal}.
Not only are visual content and images important to study, visual content 
also provides an opportunity to index and track similar content, more so than text~\cite{dubey2018memesequencer,moreira2018image,zannettou2018origins}. 

Given their influence, images naturally had major impact on politics and political campaigns~\cite{corrigall2012power,zakem2018exploring,casas2019images,bhargava2020mapping,bebic2018not, marchal2021investigating}.
For example, Casas and Williams argue that images support mobilizing because they trigger stronger emotional reactions than text, and shown that images on Twitter have a positive mobilizing effect to get people involved in online protests~\cite{casas2019images}. 
The use of images in a political context is deliberate. For example, past work on Media Cloud had shown diverging visual narratives used by media sources of different political affinity~\cite{bhargava2020mapping}.
In social media context, Marchal et al. investigated the visual formats and content themes of images shared on Twitter during the 2019 EU parliamentary election campaign, showing that both photographic and graphic (including text quotes) images were commonly shared during this campaign~\cite{marchal2021investigating}.

Other works focused specifically on the ``meme'' visual format (i.e. captioned pictures that tend to spread widely online)~\cite{miltner2018internet,gal2016gets}, with or without political context~\cite{ling2021dissecting,bebic2018not,zakem2018exploring,howley2016have,dubey2018memesequencer,zannettou2018origins,dang2017offline,vickery2014curious}.
For example, Zakem et al.~\cite{zakem2018exploring}
reported several famous meme examples and examined how these memetic engagements have been used in U.S. government influence campaigns, 
indicating the wide range of these memes in terms of content, format, text sentiment, and their role. 
For example, the authors showed how these memes were utilized by various online actors including governments and individuals. The memes often transcended individual cultures and languages, and reached broad communities. %
Beyond the political and campaign context, research had investigated the visual elements that distinguish image memes that are highly viral on social media, by developing a codebook to characterize the memes, and using it to annotate 100 memes collected from 4chan’s Politically Incorrect Board (/pol/)~\cite{ling2021dissecting}. 
The results showed that highly viral memes are more likely to contain characters or positive/negative emotions, and that memes presenting long text or without a clear subject are unlikely to be re-shared.

Naturally, and especially serving as a focus of attention in recent years, images have been used for misinformation and manipulation~\cite{firstdraftInformationDisorder,marwick2017media,paris2019deepfakes}.
The field known as visual misinformation examines how images are used for creating and propagating misinformation, while leveraging the persuasive power of images and their ease of spread~\cite{zannettou2018origins,highfield2016instagrammatics,moreira2018image,zannettou2020characterizing}.
Visual misinformation can be expressed in spreading visual fabricated or manipulated content (e.g. photoshopped content, deep fakes, etc.)~\cite{paris2019deepfakes,marra2018detection,marra2019gans,chesney2019deep}, or in sharing genuine images but out of their real context (e.g. reuse an old image and share it as new)~\cite{OutofContextPost,firstdraftInformationDisorder}. 
Various papers aim to compare these two types of visual misinformation and their spread, and develop techniques
to detect visual misinformation~\cite{OutofContextPost,matatov2022dataset,matatovdejavu,moreira2018image,gupta2013faking}.
Of course, images are often used for misinformation in the political context and in political campaigns~\cite{types_election_misinformation}. 
Indeed, the topic of the spread of misinformation during election campaigns, including the usage of visual misinformation, has received a lot of attention in recent years, which we review and expand on more broadly next.

\subsection{Election Misinformation}
\label{subsec:election_background}

A number of studies have specifically focused on elections in recent years, studying and documenting the spread of misinformation around elections and election-related misinformation campaigns (e.g.,~\cite{abilov2021voterfraud2020}), as well as seeking to understand the impact of misinformation on the outcomes and confidence in elections (e.g., ~\cite{Berlinski}).
Many of the studies looked at data from different social media platforms, understanding how content and misinformation was spread on these  platforms~\cite{grinberg2019fake,allcott2017social,rizoiu2018debatenight,vitak2011s,CenterforanInformedPublic2021,ferrara2020characterizing,abilov2021voterfraud2020,bovet2019influence}.
Particularly, several papers were published regarding the 2016 U.S. elections, studying misinformation spread on Twitter~\cite{grinberg2019fake,bovet2019influence} to understand the exposure to misinformation by voters. 
For example, Grinberg et al.~\cite{grinberg2019fake} showed how only 1\% of users were exposed to 80\% of fake news, and that an even smaller percentage of users were responsible for sharing most of it. 
Other work explored the cross-platform social media influence and misinformation by the Russian Internet Research Agency (``Russian trolls'') during the campaign~\cite{arif2018acting,golovchenko2020cross}. %

Research attention continued during the 2020 U.S. elections,
where the most prominent misinformation campaign
was  Stop the Steal. 
In this campaign, claims regarding voter fraud were spread on different media and various social media platforms~\cite{CenterforanInformedPublic2021,superspreader,murdockmulti}. 
The promoters of voter fraud claims shared narratives, images and URLs in support of (mostly false) allegations~\cite{NYTelectionMisinformation}, including missing ballots, ballots cast by dead voters, voting machines irregularities like vote switching, and other claims about the voting process. These claims served as backdrop for pressuring government officials to overturn the results~\cite{WPvoterfraudAllegation,FindVotesGeorgia,nytimestweets,callElectionCorrupt}. The misinformation campaign also targeted the topic of mail-in voting~\cite{benkler2020mail,NYTFalsehoodsMailIn}.
The Election Integrity Partnership’s election report ~\cite{CenterforanInformedPublic2021} extensively documents election-related misinformation narratives and the cross-platform transmission between a wide range of social media platforms during the 2020 election ``voter fraud'' misinformation campaign. 
The report demonstrated the participatory nature of the misinformation campaign, showing wide participation of both low- and high-level actors (e.g. verified accounts) in spreading the narratives, while leveraging the features of each social media platform to spread content effectively.

A number of research efforts had collected and made available social media data from this campaign, including content, narratives, URLs, and images~\cite{abilov2021voterfraud2020, chen2020election2020}.
In the VoterFraud2020 dataset paper~\cite{abilov2021voterfraud2020}, the authors perform an analysis of the Twitter users who shared the claims. 
Based on the analysis, the dataset labels a subset of the users as promoters or detractors of voter fraud claims, based on their participation in the retweets graph. 
The authors also presented an initial exploration of the content, including the shared images.
The researchers showed, for example, that the most popular images on Twitter were often shared in a small number of original tweets that were then retweeted many times. 
We use this dataset and the users' labels in our work here. 
Other work about the 2020 campaign built on this dataset. For example, one study used shared URL data to detect polarized
communities on Twitter~\cite{nair2021polarized}, and showed that the polarization around the fraud claims on Twitter is evident without considering tweet content other than URLs. 
Another work, using a multi-platform dataset, used URLs to compare communications about voter fraud claims within several social media platforms, and found that social media content about election fraud differ in content and also timing across platforms~\cite{murdockmulti}.

Our work here aims to specifically examine what \textit{visual content} practices are used during misinformation campaigns, by using the Stop the Steal campaign of the 2020 U.S. elections as a case study.
While previous work~\cite{abilov2021voterfraud2020} presented a basic exploration of the images used in this campaign, there is a gap in understanding the type of images used, their characteristics, the role they played in the campaign, and the overall dynamics of their spread as we do in this work.

\section{Dataset}
\label{sec:dataset}

This work uses the case study of the Stop the Steal campaign surrounding the 2020 U.S. election, focusing on the most popular images shared on Twitter as part of this campaign. 
To identify and study these images, we use the VoterFraud2020 dataset~\cite{abilov2021voterfraud2020}, a multi-modal Twitter dataset with tweets and retweets that includes key phrases and hashtags related to voter fraud claims surrounding the U.S. 2020 election, collected between October 23rd and December 16th, 2020.
The VoterFraud2020 dataset includes 7.6M tweets and 25.6M retweets from 2.6 million unique users.
The dataset includes an analysis that is able to categorize 73.8\% 
of the users in the data into two groups: %
promoters of voter fraud claims or detractors of these claims.

The VoterFraud2020 dataset also contains the perceptual hash (pHash) values 
of the 167,696 media items -- images and videos -- that were shared in the tweets.
Amongst them, 109,310 are unique pHash values. 
Perceptual hash values are binary strings designed such that the two hashes are identical when the two corresponding images perceptually seem identical. 
We used the pHash values~\cite{monga2006perceptual} %
for matching near-duplicate images shared in this dataset, i.e., finding all appearances of images across different tweets in the dataset.
Matching pHash values is known to be effective in finding repetitions of the \textit{same} image~\cite{zauner2011rihamark}.

Using the pHash values and matches, we extracted the most popular images by promoters of voter fraud claims based on the VoterFraud2020 labels~\cite{abilov2021voterfraud2020}. 
The popularity is defined as the number of shares (tweets and retweets) of the images across all tweets where they appeared, matched using the pHash value.
\newt{We aim to specifically examine images that became highly popular in the campaign. We therefore analyzed only images that were shared at least 1,000 times each within the dataset, resulting in a set of 40 images.
}
\newt{To examine the consistency and representativeness of these top forty most retweeted images, we also reviewed the following twenty images, and observed similar trends. We further discuss the representativeness of the top forty images in Section~\ref{discussion}.}
The top forty images are attached to this submission as a supplementary file\newt{, which also includes the twenty additional images we explored for validation.}

We provide further details and metadata about the images in the dataset, showing that the coverage is substantial but still includes some gaps. 
Details about the images are included in Table~\ref{table:top_40_images} in Appendix~\ref{appendix:images},  
summarizing the popularity of the images, as well as the total number of tweets and retweets in the dataset by promoters of the claims and others. 
Note that even the most retweeted images appeared in just a few different original tweets in the dataset.
For example, the image we refer to as most popular within the promoters cluster indeed was retweeted 10,424 times by promoters but only 20 times by detractors in our dataset. The image appeared in only eleven different original tweets in the dataset.
As the original work presenting the dataset points out, the dataset is estimated to cover over 60\% of the
content shared on Twitter using the selected voter fraud keywords~\cite{abilov2021voterfraud2020}.
This gap in coverage explains the difference between the number of image shares within the dataset and the total number of shares according to the Twitter-provided metadata (as of December 16th, 2020), as shown in Table~\ref{table:top_40_images}.
For example, according to metadata, the tweets containing the most popular image referenced above were retweeted a total of 20,104 times. 
Furthermore, as noted above, 26.2\% of the users in the VoterFraud2020 dataset did not have promoters or detractors labels~\cite{abilov2021voterfraud2020}, and thus the ``Total Retweets in the Dataset'' column may have higher values than the sum of the ``Retweets by Promoters'' and ``Retweets by Detractors'' columns in Table~\ref{table:top_40_images}.

\newt{We validated the effectiveness of matching exact pHash values for our task, including an exploration of the impact of using a different match threshold. 
As we show next, the matching in general proved to be effective in aggregating appearances of the same image, and false positives and negatives had minimal impact on the outcome.}
One false negative resulted in a single pair of almost-identical top images which had different pHash values (images ranked as nine and ten); we did not combine these images to maintain consistency in applying the matching criteria.
\newt{To assess whether we are missing a set of matching near-duplicates and modified images that do not have exactly identical pHash values, we ran the same aggregation process, but grouping together images with similar (i.e., non-identical) hash values, using a distance threshold. Drmic et al.~\cite{drmic2017evaluating} evaluated the robustness of different perceptual image hashing algorithms, and different distance parameters. The authors showed that even for quite aggressive image modifications, the pHash algorithm was successful, when used with a distance threshold equal to~14 (hash values in a binary representation, with a hash size of 64 bits; the distance is computed using Hamming distance).
We followed the conclusions of Drmic et al. to group together images with pHash values with Hamming distance of 14~bits or lower.
Even under this grouping, the~37 most popular images are the same as the top 37~images (out of the top 40 images) in the exact-match approach. 
Further, considering near-duplicates, the top ten images did not change their ranking at all compared to the original list; and among the following 27 images (11-37), only three images had a different relative rank (shifting up between one to eight places). 
We conducted additional analyses to verify that not only the top forty images are not replaced by others, but that we are also not missing a significant number of shares of the original top forty images because of the strict matching. 
We used the near-duplicate matching approach, with the threshold of 14 bits, to find near-duplicates of the original list of top forty images. 
We observe that considering these near-duplicate matches only adds 15.5 shares per image on average -- excluding a single image with an increase of additional 1,604 shares. 
This exception, however, can be considered a false positive match: the original image is a screenshot of a tweet, and the pHash-based near-duplicate match for this image resulted in several screenshots of \textit{other, different} tweets. 
Overall, then, in our analysis, matching exact pHash values resulted in a low false positive rate, and at most only resulted in a minor difference in rank and share counts, justifying the exact-match approach described above.}
\newt{Our main unit of analysis is therefore an image, where we consider all images that have the same unique hash value as the same image.}

For each examined image, we explored the tweets and retweets where the image was shared.
For each share, i.e. tweet or retweet, the VoterFraud2020 dataset provides extensive metadata, including the timestamp of the post, the user who posted it, the user's cluster (detractor or promoter of voter fraud claims), and the number of followers the user has.
Nevertheless, the user's detailed metadata in the \newt{VoterFraud2020} dataset is limited to users who posted an original tweet at least once within the dataset, as well as users that only retweeted within the dataset but remained active on February 1st 2021, when their data was retroactively collected. The metadata of the users who posted original tweets was collected along with user's first share in the dataset, and is timestamped in the original VoterFraud2020 dataset~\cite{abilov2021voterfraud2020}. 
In our analysis, we found that for each analyzed image, fewer than 10\% of its shares had missing user metadata (the average is approximately 7\%).
Due to these gaps, in all further analyses regarding the number of followers and image exposure we ignore shares with missing user metadata.
For all other analyses (e.g. temporal patterns of spread), we considered all shares. 

As noted above, some of the media items in the VoterFraud2020 dataset were actually videos posted along with the tweets, appearing as media in the tweet metadata.
In order to focus on images only, we viewed the top tweeted media items, using the Internet Archive\footnote{https://web.archive.org/} when the original tweet was not available, and skipped media items that were videos. \newt{There were a total of~31 videos in the dataset that exceeded our threshold of shares, and their popularity was interleaved with that of the 40 images in the data.}

\section{Analysis of Image Categories}
\label{sec:categories}

Using the dataset described in Section~\ref{sec:dataset}, we examined the forty most popular images,
first taking a qualitative approach for categorizing and characterizing these images. 
This examination was done to understand the types of images that reached popularity in the campaign, and their role in it. 
Moreover, in the next sections (sections~\ref{sec:temporal} and~\ref{sec:users}), we use some of these labels to analyze the different patterns of sharing and spread of these images. 

\subsection{Creating a Codebook for the Images}
\label{subsec:categories:creating}
To perform the qualitative analysis, we used the images as well as the context of how the images were shared in tweets.
For each image, we examined the image itself and up to ten tweets that shared the image, looking at the tweets' text and associated URLs if any. As Table~\ref{table:top_40_images} shows, some images were shared in fewer than 10 original tweets, in which case we reviewed all of them. For images shared in more than ten original tweets, we examined the first five tweets in our dataset that shared the image, and the most recent five tweets.

\newt{The method and multi-label approach of generating the codebook was inspired by previous work aiming to characterize similar content, such as \newt{images~\cite{marchal2021investigating},} memes~\cite{ling2021dissecting} or posts~\cite{lopez2013consequences}.
Similarly to these papers, we developed the codebook using an iterative process composed of multiple steps. 
First, we examined the image dataset, identified the images' most noticeable characteristics, and assigned these emergent labels to the images. 
This initial step was carried out by the first author, and the next steps were performed by the entire research team.
Based on the initial assignment of multiple labels for each image, we organized the labels into a  higher-level set of categories, i.e., different dimensions that characterize the image. 
After defining the categories, we added more labels to make them more complete, and then iterated on redefining the categories, and so on.
We repeated these steps, discussed and refined, until the codebook reached stability and agreement of the research team.
The next section presents the categories and labels, their interpretation, and their assignment to the images.}

\subsection{Categories and Labels}
\label{subsec:categories:outcome}
Our labeling procedure exposed the different types and uses of the popular images in the dataset, represented in a set of categories and labels we describe in detail below.
The categories and their associated labels are summarized in the tables below, and include the categories Type (Table~\ref{table:type_tags_number_of_images}), Origin (Table~\ref{table:origin_tags_number_of_images}), Content (Table~\ref{table:content_tags_number_of_images}), and Role (Table~\ref{table:role_tags_number_of_images}).
The tables also include a description for each of the labels, along with the number of images that were assigned the label among the forty images we used as our data. 
Note that for some categories, noted below, multiple labels can be assigned to each image. For other categories, the labels are mutually exclusive. 

For additional illustration, Table~\ref{table:example_images_labels} shows two of the images and their complete set of labels in each category.
Image \#2 is a photograph presenting allegedly buried ballots that in some cases was tweeted along with links to articles presenting the image and text saying that this image is evidence of voter fraud. 
On the other hand, image \#9 includes text, rendered visually as a poster, and serving as a call-to-action while indirectly claiming the existence of voter fraud.

The labels of the top forty images are attached to this paper as a supplementary file along with the images and their pHash values. %
Using these pHash values, our work is reproducible when used in conjunction with the VoterFraud2020 dataset~\cite{abilov2021voterfraud2020}. 
The pHash values can be used across datasets, allowing researchers to expand the work (and the application of the labels) beyond Twitter. 

\begin{table}[t]
\footnotesize 
\begin{tabular}{|c|c|c|}
\hline
\textbf{\begin{tabular}[c]{@{}c@{}}The Image\end{tabular}} & 
\includegraphics[height = 1.1in]{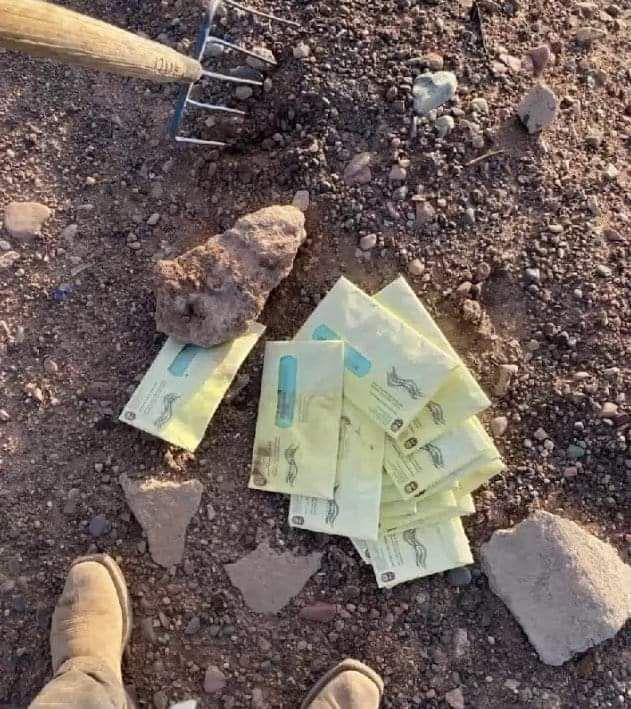} & 
\includegraphics[height = 1.1in, width=1.37in]{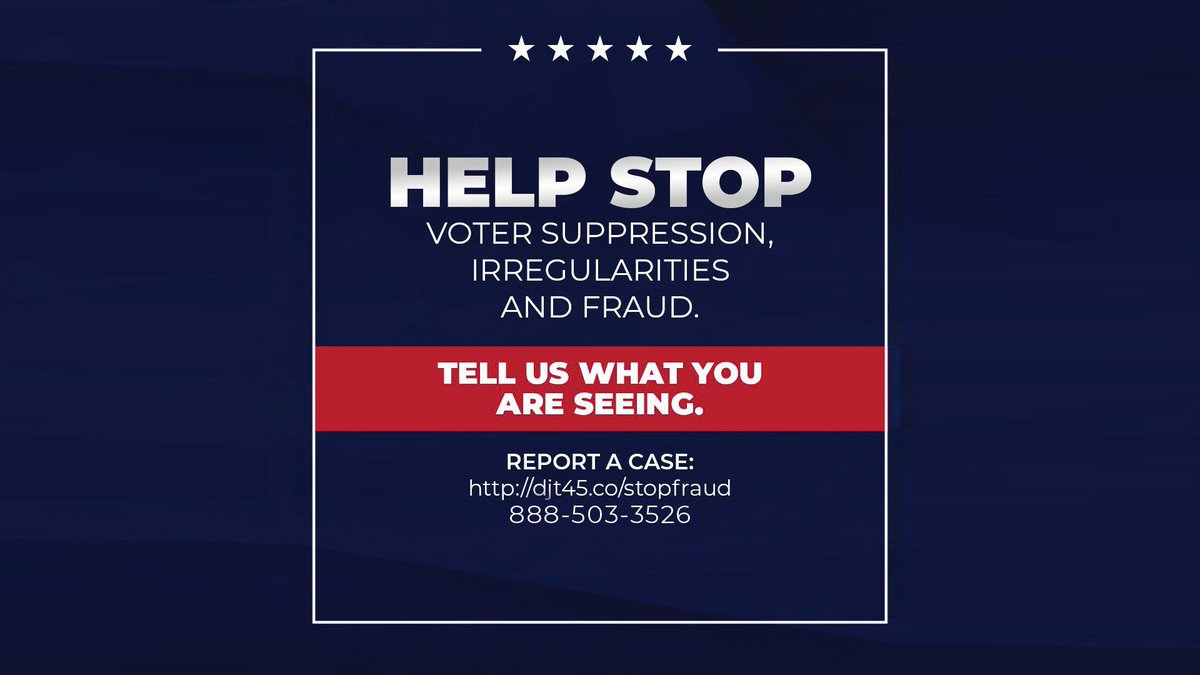} 
\textbf{} \\ \hline
\textbf{\begin{tabular}[c]{@{}c@{}}Popularity\\ Rank\end{tabular}} & 2 & 9 \\ \hline
\textbf{Type} & Photograph & Text as image \\ \hline
\textbf{Origin} & \begin{tabular}[c]{@{}c@{}}Online media article\end{tabular} & \begin{tabular}[c]{@{}c@{}}Computer-generated\end{tabular} \\ \hline
\textbf{Content} & \begin{tabular}[c]{@{}c@{}}Photographic image of voting\end{tabular} & Poster \\ \hline
\textbf{Role} & Evidence & \begin{tabular}[c]{@{}c@{}}Claim of voter fraud,\\ Call to action\end{tabular} \\ \hline
\end{tabular}
\caption{Two examples of images from the dataset, and their labels from each category.}
\label{table:example_images_labels}
\end{table}

\paragraph{\textbf{Type.}}
This category captures the type of data presented in the image, with each image having exactly one Type label. 
Table~\ref{table:type_tags_number_of_images} presents the labels, sorted by the number of images that were assigned to each label.
These labels were developed separately, but echo closely the classification used by Marchal et al. for Twitter images shared in the 2019 EU election campaign~\cite{marchal2021investigating}.

\begin{table}[t]
\centering
\small
\begin{tabular}{|llc|}
\hline
\multicolumn{3}{|l|}{\textbf{Type: what is the type of data presented in the image?}}                            \\ \hline
\multicolumn{1}{|l|}{Photograph}                     & \multicolumn{1}{l|}{A photograph, visual image produced with a camera.}                                & 20 \\ \hline
\multicolumn{1}{|l|}{Text as image}                       & \multicolumn{1}{l|}{A rendering of text in visual format.
}  & 16 \\ \hline
\multicolumn{1}{|l|}{Graphic/Infographic}                & \multicolumn{1}{l|}{Information presented in a graphical form, such as charts and graphs.}                                     & 3  \\ \hline
\multicolumn{1}{|l|}{Meme}                & \multicolumn{1}{l|}{An image with text inserted over it in order to convey a message.}    & 1  \\ \hline
\end{tabular}
\caption{Labels from Type category and their prevalence within the popular images
} 
\label{table:type_tags_number_of_images}
\vspace{-5mm}
\end{table}

As shown in the table, exactly half of the most popular images we examined were photographs -- i.e., what we would consider a regular visual image, produced with a camera. More surprisingly, 40\% of the images were actually text rendered as images or some other capture of content that is mostly text (more on content below). 
Of the top ten images in our data, seven were photographs, and three were text-as-image.
A possible explanation for the large fraction of text-as-image visual content is that Twitter users used textual images as a way to attract visual attention~\cite{counts2011taking,hsieh2011different} or to increase  believability~\cite{wittenberg2021minimal}.  
Notably, research on visual misinformation often focused on memes, deep fakes, photoshopped images, etc.~\cite{paris2019deepfakes,miltner2018internet,zakem2018exploring,chesney2019deep,gal2016gets,ling2021dissecting,marra2018detection,marra2019gans}. While some images in our data include possibly-misleading imagery, none of the analyzed images were labeled as a manipulated photograph, and only one of the popular images was a photograph that was clearly edited (as a meme)~\cite{miltner2018internet,gal2016gets}.

\paragraph{\textbf{Origin.}}
The origin category captures the potential source of the image\newt{, to the extent it can be determined (as we explain below)}. For this category, each image has exactly one label. 
Table~\ref{table:origin_tags_number_of_images} presents the labels, sorted by the number of images that were assigned to each label, and split into three themed sections.

\begin{table}[H]
\centering
\small
\begin{tabular}{|lllc|}
\hline
\multicolumn{4}{|l|}{\textbf{Origin: where did the data presented in the image come from?}} \\ \hline
\multicolumn{1}{|l|}{\multirow{2}{*}{\rotatebox[origin=c]{0}{\begin{tabular}[c]{@{}l@{}}Media \\ source \end{tabular}}}}  &
\multicolumn{1}{l|}{Online media article}                   & \multicolumn{1}{l|}{\begin{tabular}[c]{@{}l@{}}An image that also appeared in URLs that were \\ attached to the tweets.\end{tabular}} & 10  \\ \cline{2-4} 
\multicolumn{1}{|l|}{}    & 
\multicolumn{1}{l|}{Media Screen Capture}                                         & \multicolumn{1}{l|}{A photograph of a TV screen.}    & 2  \\ \hline \hline
\multicolumn{1}{|l|}{\multirow{2}{*}{\rotatebox[origin=c]{0}{\begin{tabular}[c]{@{}l@{}}Individual \\ media\end{tabular}}}}  &
\multicolumn{1}{l|}{Document screenshot}                                                  & \multicolumn{1}{l|}{Screenshot of a document or an article.}   & 6  \\ \cline{2-4}
\multicolumn{1}{|l|}{}                         & \multicolumn{1}{l|}{Social media segment capture}   & \multicolumn{1}{l|}{\begin{tabular}[c]{@{}l@{}}An image of a social media screen, post, or comment \\ (Twitter or other platform).\end{tabular}} & 4  \\ \hline \hline
\multicolumn{1}{|l|}{\multirow{3}{*}{\rotatebox[origin=c]{0}{Other}}}  &
\multicolumn{1}{l|}{Computer-generated}                                         & \multicolumn{1}{l|}{\begin{tabular}[c]{@{}l@{}}Computer-generated content \\ (e.g. poster, logo, infographic, or meme).\end{tabular}}    & 8 \\ \cline{2-4}
\multicolumn{1}{|l|}{}                         & 
\multicolumn{1}{l|}{Snapshot}                                                  & \multicolumn{1}{l|}{A photograph from non-professional source.}   & 6  \\ \cline{2-4}
\multicolumn{1}{|l|}{}                         & \multicolumn{1}{l|}{Professional stock image}   & \multicolumn{1}{l|}{\begin{tabular}[c]{@{}l@{}}A photograph of a public figure that was likely\\ taken by professionals.\end{tabular}} & 4  \\ \hline 
\end{tabular}
\caption{Labels from Origin category and their prevalence within the popular images.
}
\label{table:origin_tags_number_of_images}
\end{table}

\newt{Almost a third (12) of the images in our set of forty popular images could be easily associated with an external media source outside of Twitter, either because these images originated in media articles that were attached to the tweets, or simply by observing that they are media screen captures from television or video channels.} 
The most prevalent among these labels is the ``online media article'' label -- i.e., an image that was published in an article (that was also  shared within the tweets in the dataset).
An additional quarter of the images also have their origins in media, but other media types which are more individual or personal, including screenshots from emails, documents, and social media platforms.
In all, the common presence of such media content within the popular images indicates how Twitter was commonly used to spread ideas from other media platforms. %
For the rest of the images, which did not originate from other media, it was harder to identify the source. Instead, we selected labels that capture the process through which the images were created in our evaluation, including snapshots, images borrowed from stock imagery, and graphics created on a computer. 
Note that images labeled as ``computer-generated'' were only infographics, memes, or textual content such as posters, i.e. not manipulated photographs (as noted above).

\paragraph{\textbf{Content.}}
This category captures the actual content of the image, attempting to broadly describe what one might see in each image in our set. %
For this category, each image has one or more labels. 
Table~\ref{table:content_tags_number_of_images} presents the labels, split into three sections: photographic content, text-based content, and other. 
The labels in each section of the table are sorted by the number of images attached to them.

A quarter of the most popular images referred to the voting process (ballots, envelopes, etc.) in some manner. The same number of images were labeled as presenting a public figure. Together, these two labels represent almost the entire subset of photographic images in the dataset. 
Perhaps unsurprisingly, the first label, photographic images of voting, was assigned to seven of the ten most popular images. %
While some of the images in this category attempt to present evidence of voter fraud, others showed generic representations of the voting process. 
For the textual images category, the most prevalent images were those rendering quotes from well-known individuals as an image, as well as images which are mostly text-based capture from articles or documents. 
These two labels represent different benefits of the use of visual medium. 
The ``quote'' images build on Twitter's rendering of tweets with images to enhance the presentation of content and make it more appealing. 
The capture from other sources, including from other social media users, is used to share cross-platform information in an efficient and visual manner.

\begin{table}[t]
\centering
\small
\begin{tabular}{|lllc|}
\hline
\multicolumn{4}{|l|}{\textbf{Content: what is being presented in the image?}} \\ \hline
\multicolumn{1}{|l|}{\multirow{4}{*}{\rotatebox[origin=c]{90}{Photographic   }}}  &
\multicolumn{1}{l|}{Photographic image of voting}                   & \multicolumn{1}{l|}{Ballots, envelopes, their transportation process, etc.}   & 10  \\ \cline{2-4} 
\multicolumn{1}{|l|}{}    & 
\multicolumn{1}{l|}{Public figure}                                         & \multicolumn{1}{l|}{A photo of a well-known person.}    & 10 \\ \cline{2-4} 
\multicolumn{1}{|l|}{}                         & 
\multicolumn{1}{l|}{Broadcast or interview capture}                                & \multicolumn{1}{l|}{\begin{tabular}[c]{@{}l@{}} A photograph showing TV (or other media source) broadcast \\ or an interview.\end{tabular}}                                 & 2 \\ \cline{2-4} 
\multicolumn{1}{|l|}{}                         & \multicolumn{1}{l|}{Protest}   & \multicolumn{1}{l|}{\begin{tabular}[c]{@{}l@{}} A photo of a collective public action such as protest, \\ demonstration.\end{tabular}}    & 1  \\ \hline \hline
\multicolumn{1}{|l|}{\multirow{6}{*}{\rotatebox[origin=c]{90}{Textual}}} &
\multicolumn{1}{l|}{Quote}                                                  & \multicolumn{1}{l|}{A quote from a known individual, or that is implied as such.}   & 5  \\ \cline{2-4}
\multicolumn{1}{|l|}{}                         & \multicolumn{1}{l|}{Text from an article or document}   & \multicolumn{1}{l|}{\begin{tabular}[c]{@{}l@{}}Part of an article or an official document.\end{tabular}} & 4  \\ \cline{2-4} 
\multicolumn{1}{|l|}{}                         & \multicolumn{1}{l|}{Text by social media users}                                                  & \multicolumn{1}{l|}{\begin{tabular}[c]{@{}l@{}}Text written by anonymous or ordinary user on social \\ media platforms.\end{tabular}}   & 3  \\ \cline{2-4}
\multicolumn{1}{|l|}{}                         & \multicolumn{1}{l|}{Poster}                                                  & \multicolumn{1}{l|}{\begin{tabular}[c]{@{}l@{}}Short information, mostly textual, designed as a poster and \\ intended to provide details or offer proposals.\end{tabular}}   & 3  \\ \cline{2-4}
\multicolumn{1}{|l|}{}                         & \multicolumn{1}{l|}{Other textual content}                      & \multicolumn{1}{l|}{\begin{tabular}[c]{@{}l@{}}Twitter’s misinformation warning.\end{tabular}}    & 1  \\  \hline \hline
\multicolumn{1}{|l|}{\multirow{1}{*}{\rotatebox[origin=c]{90}{}}}  &
\multicolumn{1}{l|}{Other content}                                         & \multicolumn{1}{l|}{Graphic/infographic or meme items.}    & 4 \\ \hline 
\end{tabular}
\caption{Labels from Content category and their prevalence within the popular images.}
\label{table:content_tags_number_of_images}
\end{table}

\paragraph{\textbf{Role.}}
This category captures the role that the image was meant to play in the campaign, according to our interpretation of the image and its context.
For this category, each image has one or more labels. Table~\ref{table:role_tags_number_of_images} presents the labels, split according to the type of image, and sorted by the severity of the information that we think the images were trying to convey.

\begin{table}[t]
\centering
\small
\begin{tabular}{|lllc|}
\hline
\multicolumn{4}{|l|}{\textbf{Role: what is the role of the image in the context of the campaign?}} \\ \hline
\multicolumn{1}{|l|}{\multirow{3}{*}{\rotatebox[origin=c]{90}{Photographic}}}  &
\multicolumn{1}{l|}{Evidence}                   & \multicolumn{1}{l|}{\begin{tabular}[c]{@{}l@{}}Provide (alleged) evidence of the existence of \\ voter fraud.\end{tabular}}   & 5  \\ \cline{2-4} 
\multicolumn{1}{|l|}{}    & 
\multicolumn{1}{l|}{\begin{tabular}[c]{@{}l@{}}Sow doubt in the election process \\ or outcome\end{tabular}}                                         & \multicolumn{1}{l|}{\begin{tabular}[c]{@{}l@{}}Visual hints or indications in support of the existence \\ of voter fraud, or images meant to sow doubt. \\ \end{tabular}}    & 11 \\ \cline{2-4} 
\multicolumn{1}{|l|}{}                         & 
 \multicolumn{1}{l|}{Illustration only}   & \multicolumn{1}{l|}{\begin{tabular}[c]{@{}l@{}}Provide a visual that is related to the text in a tweet, \\ but is not adding information,  e.g. relevant stock photo.\end{tabular}}    & 8  \\ \hline \hline
\multicolumn{1}{|l|}{\multirow{3}{*}{\rotatebox[origin=c]{90}{Textual}}} &
\multicolumn{1}{l|}{Claim of voter fraud}                                                  & \multicolumn{1}{l|}{Claims the existence of voter fraud in the election.}   & 13  \\ \cline{2-4}
\multicolumn{1}{|l|}{}                         & \multicolumn{1}{l|}{Call to action}   & \multicolumn{1}{l|}{\begin{tabular}[c]{@{}l@{}}Direct call to act in some way to aid the campaign.\end{tabular}} & 3  \\ \cline{2-4} 
\multicolumn{1}{|l|}{}                         & \multicolumn{1}{l|}{React against detractors of the claims}                      & \multicolumn{1}{l|}{\begin{tabular}[c]{@{}l@{}}Images that include detraction or suppression of voter \\ fraud claims, usually with tweet text that confronts it.\end{tabular}}    & 2  \\  \hline 
\end{tabular}
\caption{Labels from Role category and their prevalence within the popular images. 
}
\label{table:role_tags_number_of_images}
\end{table}

In this category, we used the ``evidence'' and the ``claim of voter fraud'' labels when the image presented a photographic or textual (respectively) attempt to provide evidence of the alleged voter fraud.
We labeled five images as ``evidence'', and 13 as ``claim of voter fraud''. 
The five images labeled as evidence present content such as lost ballots, discarded envelopes, and discarded ballots boxes.
While only a few such images emerged, these images were some of the most commonly shared: four of these ``evidence'' images were included in the ten most shared images, and the fifth was ranked at 15.
While the ``evidence'' photographs were some of the most popular images in this dataset, the most common label assigned to images in our dataset was the ``claims of voter fraud'' label for text-based images. 
These 13~images included textual claims that voter fraud occurred, rendered as an image to increase visibility and aid spread. 
For example, personal stories that individuals shared about how they or their acquaintances were witness to voter fraud, or alleged legal and official documents about the subject (e.g., media advisory notice of hearing about election integrity).

Eleven additional images presented visual hints as indirect photographic indication of the idea of fraud, by using snapshots related to the subject that are not necessarily presented as ``evidence'', for example photos of arrests allegedly due to fraud, or suspicious scenes from the voting process that presented to sow doubt.
These images do not provide any actual evidence such as photograph of stolen ballots, but rather present photographs that, along with the tweets' text, encourage suspicion. 
The most popular image in the dataset, as well as the third most-shared image, were both assigned this label. The images showed moments from the process of scanning or transferring ballots, and the tweets accompanying them included text claiming wide-spread fraud.  %

Finally, images were also used in tweets that claimed a disorder related to voting, but without having a direct photograph from the event, shared an illustrative image, usually with content related to the claim. For example, one popular image (\#18) included a photo of the person that was mentioned in the text. We labeled eight images as illustrations, and none of them were among the 10 most popular images.

Other labels in the textual type of image were less prevalent. Three images presented text that directly calls to action, for example, a poster calling to report on fraud events. This direct usage of images to motivate actions aligns with prior work that shows that images have a positive mobilizing effect in the context of online protest activity~\cite{casas2019images}. Two anomalous images presented suppression of voter fraud claims, for example, a screenshot of an article headline that denies the alleged fraud. These two images were probably included in our image dataset because of their context -- they were published along with tweets that confronted the text in the images.

\section{Analysis of The Temporal Patterns}
\label{sec:temporal}

Building on the image dataset and labels described in sections~\ref{sec:dataset} and~\ref{sec:categories} we can perform new analysis to better understand the spread of these popular images over time, identify typical patterns (if any), and explore whether the temporal patterns are related to the images' labels.
In this section, we focus on the temporal dimension in the sharing of the images by Twitter users.

\begin{figure*}
    \centering
    \includegraphics[width=\textwidth]{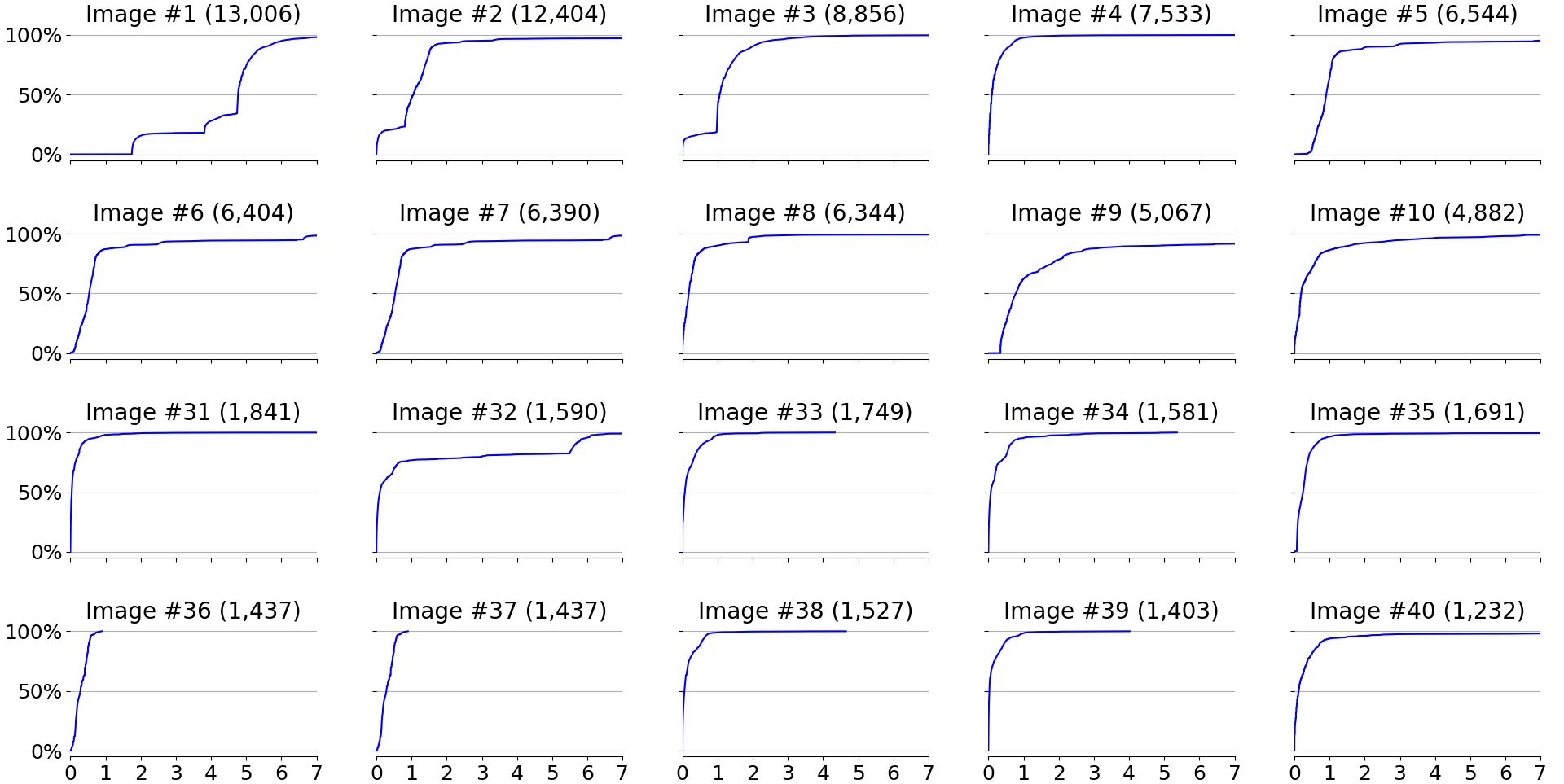}
    \caption{
    Cumulative graphs showing, for a subset of our images, what proportion of its total shares (Y-axis) each image received over the first seven days (X-axis) since it first appeared in our dataset.  
    Above each graph is the index of the image and the total number of shares the image had in our the dataset.  
    }
    \label{fig:cumulative_shares_7_days}
\end{figure*}

We study the spread over time of tweets and retweets that included the images.
Figure~\ref{fig:cumulative_shares_7_days} presents cumulative graphs of the tweets and retweets which shared each image over time, for the images ranked 1--10 and 31--40 in popularity based on shares in our data (the figures for images 11--30 were omitted because of space considerations, but presented similar patterns). 
These plots show the cumulative proportion of the total shares of the image in our dataset since its first appearance (Y-axis), over the time since it was first shared (X-axis), limited to the first seven days since the image appeared in the data.
For example, the most popular image (top left) did not reach even half of its total eventual shares until after five days since its first appearance.
However, as shown in the figure, most of the images peaked very quickly after their first appearance in the dataset. 
Indeed, most images reached close to 100\% of their shares within the first few days, and often within the first day (see, for example, image~\#4).
While several images -- most notably, the top three images -- had more than one significant period of rapid spread, most had only one period of rapid distribution. 
We did not discern any significant and consistent differences in the temporal patterns of how the images were shared depending on their category labels, or according to their relative rank within the forty most popular images.

\begin{figure*}
    \centering
    \includegraphics[width=1.0\textwidth]{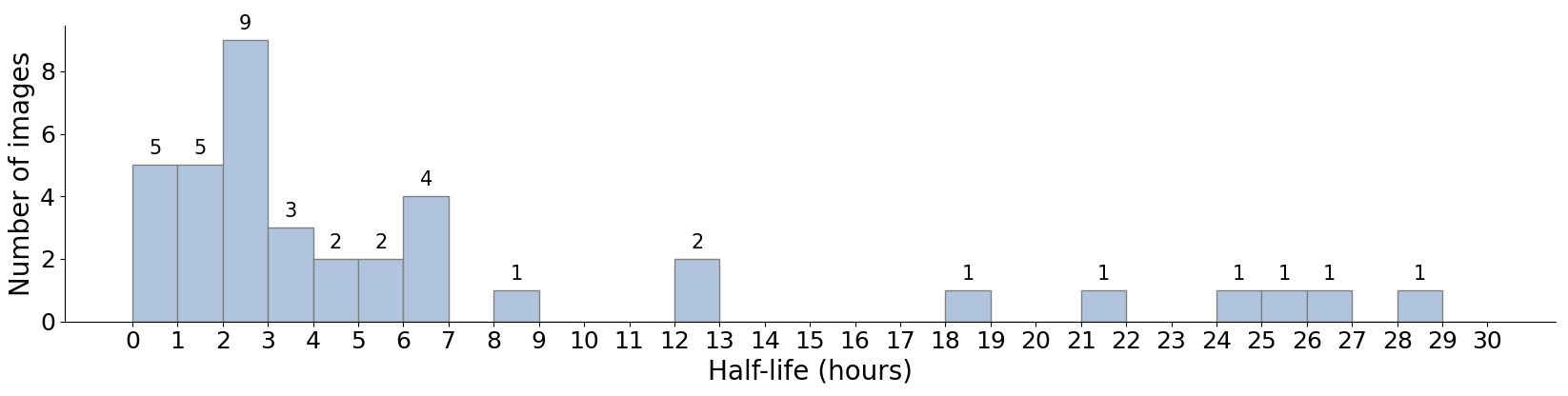}
    \vspace{-7mm}
    \caption{Histogram of the time (in hours) between each image's first appearance until it got half of their total shares in the dataset.}
    \label{fig:two_half_times_hist}
\end{figure*}

To better quantify the speed in which these popular images were spread, we calculated the half-life of the spread for each image, i.e., the time until each image reached half of its total shares in the dataset.
Figure~\ref{fig:two_half_times_hist}
presents a histogram of the half-life values for all forty images, with the X-axis bins, in hours, representing the time from the first share of the image in our data until it reached its half-life. The Y-axis shows the number of images that reached their half-life in this time frame. 
Note that the histogram excludes image \#1 (the most popular image), as its half-life was 114.4 hours. %
The histogram shows that a total of 19 images reached their half-life within three hours (the sum of the left-most three columns). After seven hours at most, 75\% of the images already reached half of their total number of shares.

To understand whether sharing of the images and actual exposure to them followed a similar spread, we also calculated the half-life of the image \textit{exposure}: the amount of time until the images got half of their cumulative followers' exposure. However, limited by the (non-)availability of network data from Twitter and in the VoterFraud2020 dataset, %
our analysis was done by simply adding up the number of followers of each user who shared the tweet (excluding a small subset of users for whom we did not have the user's metadata, as noted in Section~\ref{sec:dataset}).
\newt{Since users can have overlapping followers, this measure provides an upper bound of the actual cumulative followers' exposure.}
The temporal patterns of exposure were very similar to the sharing half-life, with most of the images reaching their exposure half-life within 3 to 5 hours.

\begin{figure*}[t]
    \centering
    \includegraphics[width=0.8\textwidth]{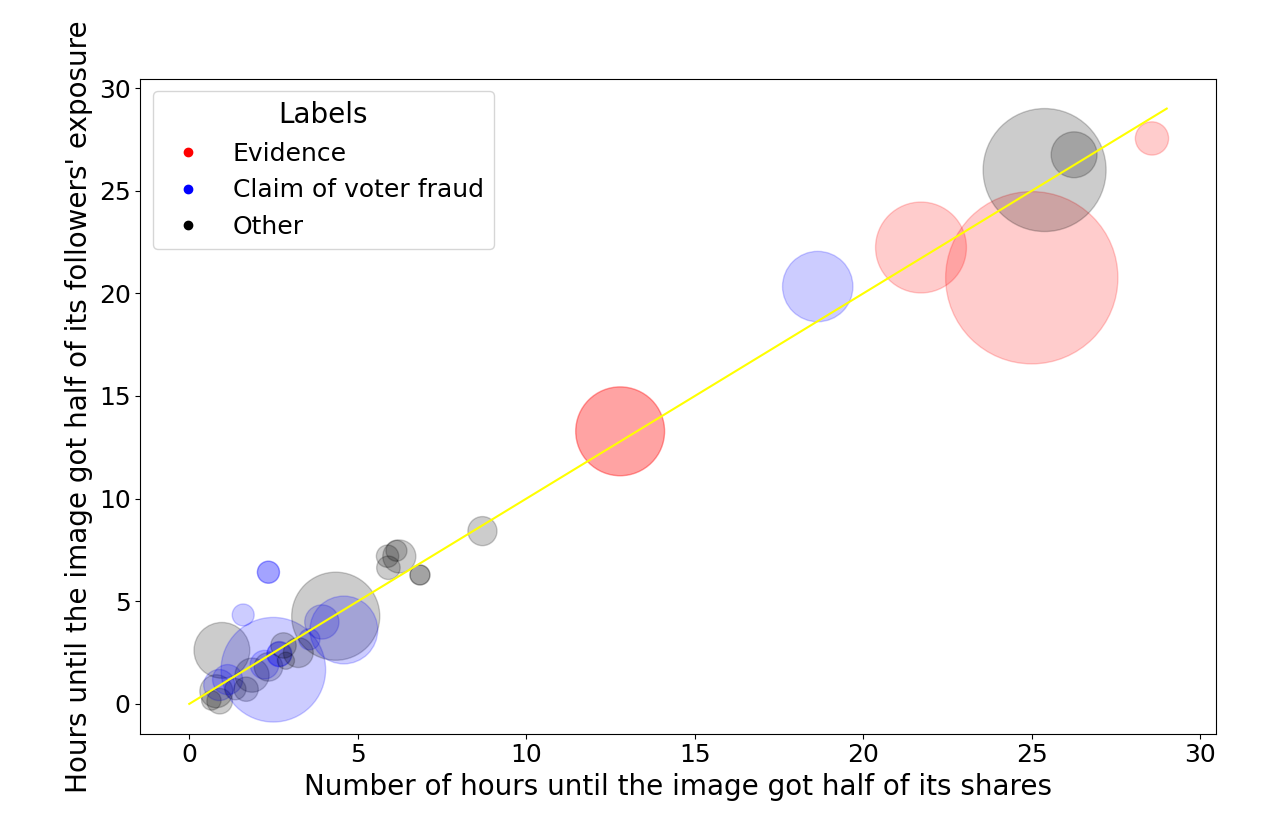}
    \vspace{-2mm}
    \caption{Comparison of the time it took for each image to reach its half-life in terms of shares (X-axis), and in terms of total followers exposure (Y-axis).
    The size of each marker is proportional to the total number of shares of the image in our dataset, and the colors represent different labels from Role category.
    }
    \label{fig:two_half_times_line_plot_role}
\end{figure*}

Figure~\ref{fig:two_half_times_line_plot_role} shows the high correlation between shares and exposure \newt{(r > .98, p < .001)}, 
and explores the difference in the patterns of sharing of images with different Role labels.
The figure shows, for each image, the half-life in terms of shares (X-axis) and in terms of followers exposure (Y-axis) (again excluding image \#1). 
The size of each marker is proportional to the total number of shares of the image in our dataset. The color of each marker corresponds to key labels from the Role category (Section~\ref{sec:categories}). 
The figure shows that for most images, the two half-life values were similar. 
Moreover, the figure shows that while some of the more popularly shared images had longer half-lives (top right of the image), this trend was not entirely consistent. 
At the same time, the labels marker colors show that the popular images that served as photographic evidence of the existence of voter fraud (``evidence'' label in Section~\ref{sec:categories}) tended to be the ones that took longer to be shared, with relatively high half-life values.
In contrast, almost all images that were labeled as text claiming voter fraud, as well as other labels, had relatively short half-lives.
We performed similar analysis to compare labels from the other categories, but those did not show any meaningful trends.

The high correlation of shares and exposure suggests that the involvement of popular- and less-popular users remained relatively stable in the different stages of the images' spread. We examine the patterns of the users participating in the spread in the next section.

\section{Analysis of the Sharing Users}
\label{sec:users}

While Section~\ref{sec:temporal} focused on the temporal patterns of the popular images, here we turn our attention to the users who shared them and their characteristics.
In particular, we examine the patterns of how popular- and less-popular users participated in sharing these images. We introduce an analysis that aims to explore whether some of these popular users were likely to participate more, or earlier, than others. \newt{Also, we include a qualitative review of the type of content shared by these more eager users.}

First, we explored the distribution of shares of the examined images, based on the popularity of users sharing them.
Previous work documented what the researchers termed multi-directional participation in the Stop the Steal campaign~\cite{CenterforanInformedPublic2021}. The term meant to capture the fact that both popular- and less-popular users participated in the spread, and the influence flowed in both directions.
Our analysis is geared to explore a similar question 
by looking at how users with different levels of popularity participated in the spread of the images in our dataset.  
As a general overview, Table~\ref{table:users_sizes} shows the size of the different accounts participating in sharing these images. 
The table shows the number of shares by each group of users, as well as the percentage of shares by this group of the total number of shares in the dataset. 
The table shows, for example, that only 16 shares were performed by users with more than one million followers, and none of the shares of the popular images in the dataset was by a user with more than ten million followers.
Moreover, most shares (88.3\%) were done by users with relatively fewer followers, i.e. users with 5,000 followers at most, and almost half the shares by users with fewer than 500 followers.
At the same time, these numbers expose a significant participation of accounts with a larger follower base, as over 5\% of the shares performed by users with more than 10,000 followers.

\begin{table}[b]
\small 
\begin{tabular}{|c|c|}
\hline
\textbf{Number of followers} & 
\textbf{\begin{tabular}[c]{@{}c@{}}Shares (and \% of shares) \\ by this group \end{tabular}} \\ \hline
0 - 50 & 19,579 (15.4\%) \\ \hline
50 - 100 & 10,463 (8.2\%) \\ \hline
100 - 500 & 32,655 (25.6\%) \\ \hline
500 - 1K & 15,299 (12\%) \\ \hline
1K - 5K & 34,452 (27.1\%) \\ \hline
5K - 10K & 7,592 (6\%) \\ \hline
10K - 100K & 7,051 (5.5\%) \\ \hline
100K - 1M & 243 (0.2\%) \\ \hline
1M - 10M & 16 (0.01\%) \\ \hline
10M+ & 0 (0\%) \\ \hline
\end{tabular}
\caption{The participation of users with different number of followers in spreading the popular images.}
\label{table:users_sizes}
\end{table}

\begin{figure*}[h]
    \centering
    \vspace{-2mm}
    \includegraphics[width=1.0\textwidth]{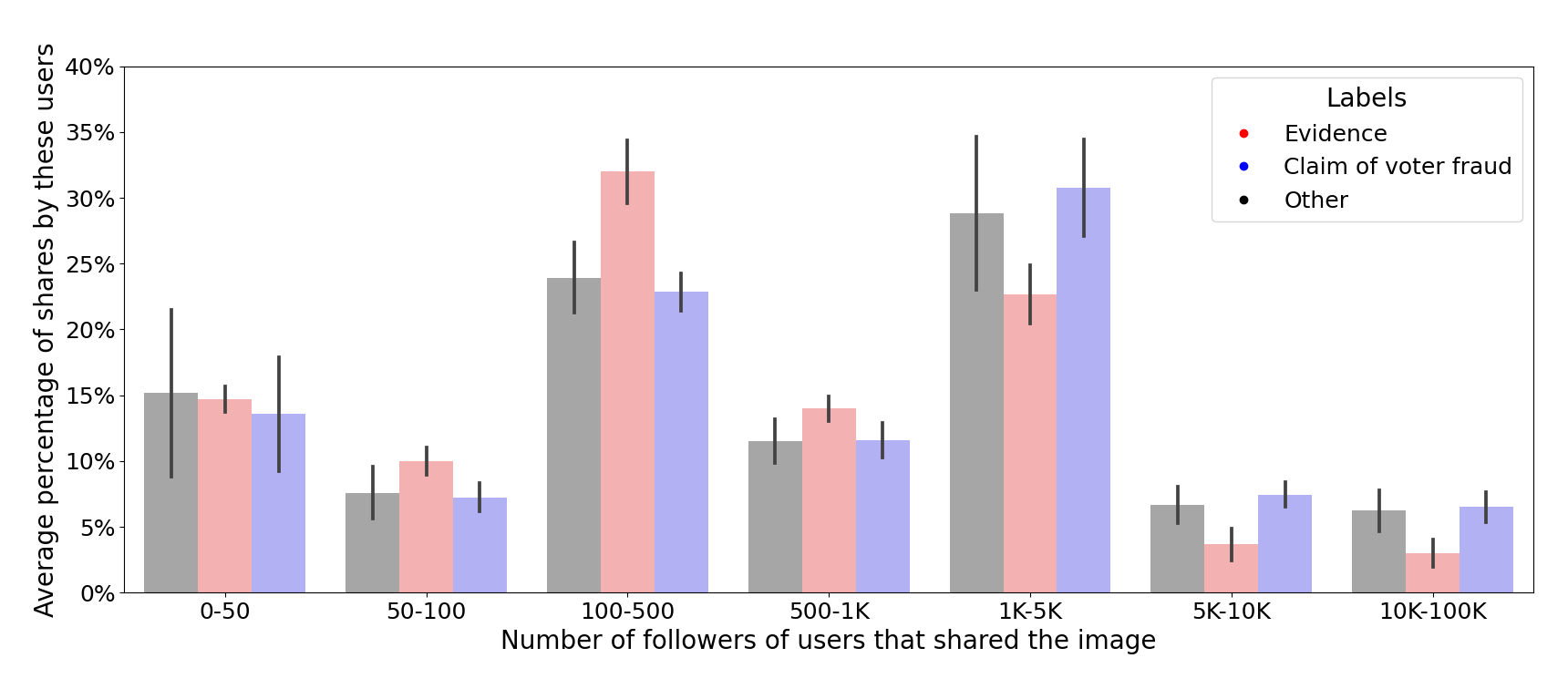}
    \vspace{-8mm}
    \caption{A comparison of the average proportion of shares of images (Y-axis) by users with different levels of popularity (number of followers; X-axis).
    The different color columns represent images that correspond to specific labels from the Role category. }
\label{fig:hist_shares_user_size_label_partial_role}
\end{figure*}

Were there differences in participation of higher-following accounts, based on the type of images that were shared?
To explore this question, we examined the distribution of shares by users with different number of followers based on the image labels of the different categories.
Figure~\ref{fig:hist_shares_user_size_label_partial_role} presents a histogram of participating users based on the major labels in the Role category: `evidence', denoting a photograph that provides (alleged) evidence of fraud; `claim of voter fraud', denoting images which contain textual claims about the existence of fraud, and `other', for the rest of the images.
On the X-axis are the bins based on ranges of the number of followers of users that shared the image.
The Y-axis captures the average percentage of the total shares of each image that was done by this group of users. 
In other words, we looked at all images with a certain label. Then, per each image, we calculated the percentage of shares of the image by users from each group, and present the average over all images with this label.
For example, users with 50 followers or fewer (left-most bin) were responsible for, on average, roughly 15\% of the shares of ``evidence'' images.
The figure shows that the distribution of users who shared the evidence images is different from users that shared the other types of images, with ``evidence'' shared more often by users with 1,000 or fewer followers, compared to the distribution for other types of images. 
To verify that these results are not due to the different popularity of the images with different labels, we repeated the analysis for the top four images from each of the three categories, which resulted in the same pattern. 
We revisit these findings and discuss what factors may have contributed to it in Section~\ref{discussion}.
The patterns of participation did not change depending on the timing of sharing, e.g. early or late in the images' spread. We looked at the participation rates during different times in the spread of the image (e.g., the first~500 and the following 500 shares). 
These sharing patterns in terms of the participation of different groups remained consistent across these trials.

Because of their smaller numbers, Figure~\ref{fig:hist_shares_user_size_label_partial_role} does not include users with more than 100,000 followers. We note, however, that in our data, none of the users with more than a million followers shared an `evidence' image, and that these images were also shared less by users with 100K-1M followers compared to other images.

\begin{figure*}[h]
    \centering
    \includegraphics[width=\textwidth]{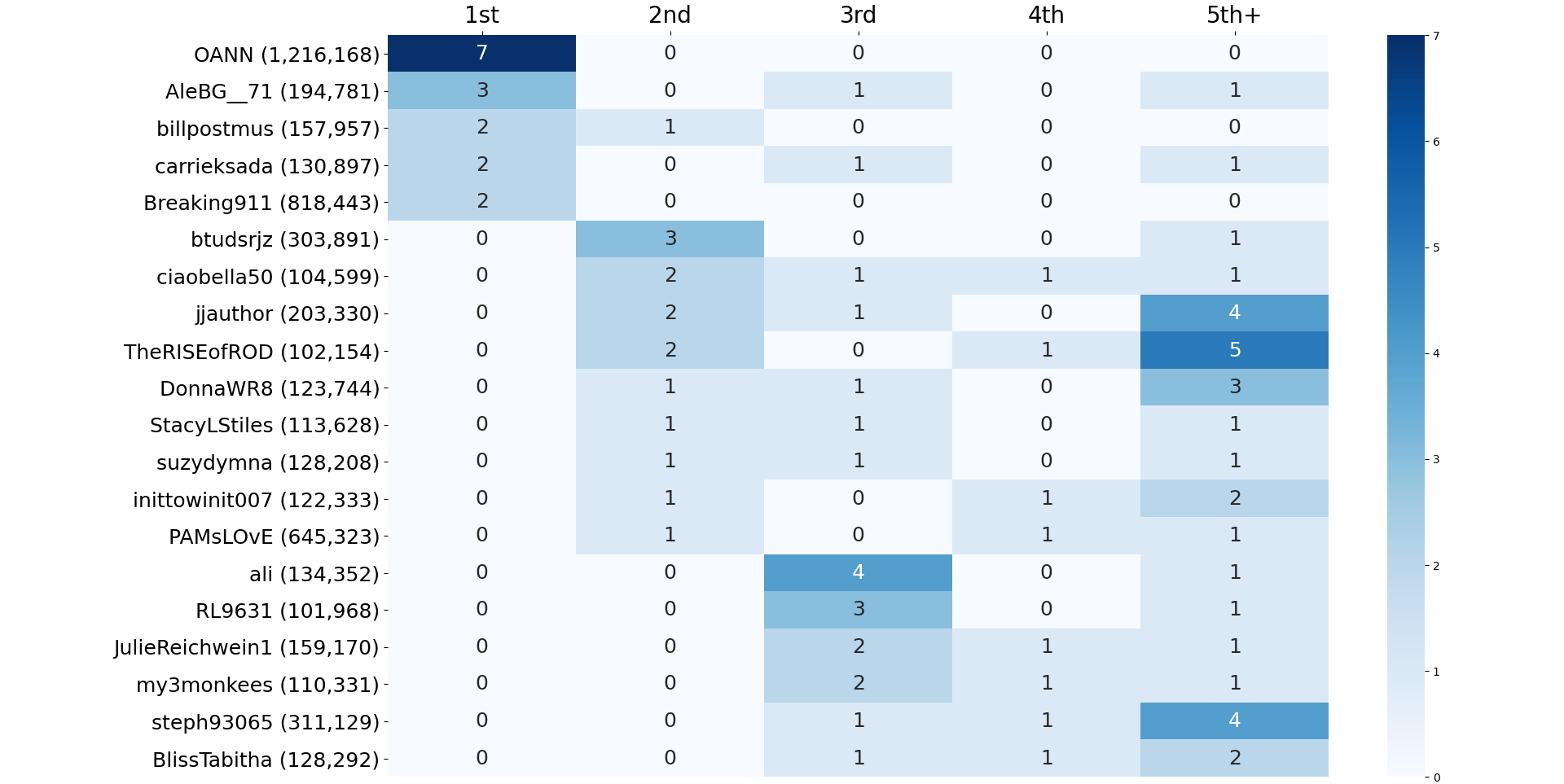}
    \caption{The earliest users, amongst those with at least 100K followers, to share images in our dataset (shows only users with at least three shares).
    For each user (row; also showing number of followers), we show how many times in our dataset they shared an image first, second, third, fourth (columns) amongst these users, and how many additional images they shared (``5th+'' column).
 }
\label{fig:fast_users_heatmap}
\end{figure*}

Nevertheless, these influential users with more than 100K followers are still important to study and understand given the impact they might have on spread of content.  
We therefore examined whether there are consistent patterns in \textit{whom} amongst the popular users participates and when.
To perform this analysis, we ran a rank-order analysis of the 146 users in our data who had more than 100K followers. These users appeared in our data (i.e. shared these top images) 258 times\newt{, where 15\% of their shares were original tweets, and the rest were retweets. All these tweets include text in addition to the tweeted image. None of these retweets in our data were quote tweets, i.e., none of them included additional text beyond the original tweet}. 
The rank in our ordering was based on each user's first instance of sharing of each image: what was their rank in sharing that image amongst the other popular users 
(e.g. 1st, 2nd, etc.). 
This ranking was then aggregated across all images in our dataset.
Figure~\ref{fig:fast_users_heatmap} shows a heat-map visualization of the results.
The figure includes the top 20 of these users amongst those who participated at least three times.
The rows are the users (represented by their Twitter handles) and the number of Twitter followers they had when the data was collected.
The columns show the frequency in which the user was ranked at this position, amongst these users, when sharing one of the images.
For example, @OANN (top row) was the first to share seven out of the images in our dataset. 
This finding is perhaps not surprising, as this account, belonging to the \newt{far-right} cable TV One America News Network\newt{~\cite{OANN_Trump_promoting,staff_behind_OANN}}, was a major participant in spreading the voter fraud claims in the media\newt{~\cite{OANN_questioning_election}}. 
These seven images the account shared were all \newt{original tweets that include text,} posted as promotion of OANN articles \newt{(all these tweets include a URL link to OANN website).
Four of the images which OANN shared have a similar pattern as they are all images presenting a photographic image of the voting process; additional two images present a public figure; and one presents a protest. 
Four of these seven images were labeled as photographic illustrations of the voting process %
(as explained in Section~\ref{subsec:categories:outcome}). 
}

More generally, Figure~\ref{fig:fast_users_heatmap} shows that amongst the popular users, a significant number were consistently ``early adopters'' in image spread. \newt{Our qualitative analysis shows that they participated in different ways. 
Beyond @OANN, the only user among the ``early and often'' users in Figure~\ref{fig:fast_users_heatmap} who posted original tweets (not only retweeted) was @Breaking911, an account which also shared tweets that include texts and URLs to articles on their own website. 
This account posted two different images, one that presented alleged evidence for voter fraud existence, and a second that aims to sow doubt in the election process or outcome}. 
Another early adopter, the right-wing activist Bill Postmus (@billpostmus), participated three times (two first-place shares amongst the top users). \newt{All his shares were retweets, that is, he did not add any text or context when sharing these images.} 
The \newt{far-right activist}
Ali Alexander (@ali), sometimes referred to as ``the man behind Stop the Steal''~\cite{Ali_Alexander_Man_Behind}, 
shared five images; for four of them he was the third of the popular users to share the image. \newt{All of Ali's shares were retweets. Four of the images he retweeted were labeled as evidence and one as call to action (Section~\ref{subsec:categories:outcome}).} 
\newt{To summarize, these ``early and often'' users shared different types of content, and participated in a mixed manner: those who posted new original tweets mostly did that to promote their own media articles; while others usually retweeted content from less-popular users.} 

Generally speaking, the results and analysis in this section provide a clearer view of the patterns of participation amongst the most popular users 
in the dataset, and provide a new way to understand the most active popular users 
in this campaign.

\section{Discussion and Conclusions}
\label{discussion}

Despite the prevalence of images in social media misinformation campaigns~\cite{firstdraftInformationDisorder,casas2019images,zakem2018exploring,OutofContextPost,types_election_misinformation,corrigall2012power,ling2021dissecting}, and a wide array of research examining visual content like images and memes in the context of social media~\cite{zannettou2018origins,highfield2016instagrammatics,zannettou2020characterizing}, research specifically focused on how images play a role in misinformation campaigns, as we do with this case study, has been limited. 
Similarly, research on election-related misinformation campaigns, including in the context of the U.S. 2020 elections~\cite{CenterforanInformedPublic2021,ferrara2020characterizing,superspreader,murdockmulti,benkler2020mail}, had not focused on the detailed analysis of the type of images and their use, as we do here.

Our analysis exposed a wide array of types and uses of images in the Stop the Steal campaign. 
Only about half of the forty most popular images were actual photographs, and many of the popular images were text-based, rendered as image for various purposes. 
Partly, these text-based images were taking advantage of the Twitter presentation of images to maximize attention~\cite{hsieh2011different,counts2011taking,wittenberg2021minimal}.
In other cases, the text-based images allowed users to bring in content from other contexts, e.g. other social media posts or documents, or to use graphic layouts to create a better call-to-action. 
Such mixed use of images on Twitter was also observed in the context of a more general political campaign~\cite{marchal2021investigating}.

Even within the photographic content, only very few of the popular photos aimed to present direct ``evidence'' of voter fraud. 
We do not know from this analysis whether there were many other such images that did not succeed in getting wide spread for some reason. 
Regardless, the evidence photographs which did make it to our dataset of top images were quite successful (most of them in the top~10 most shared images). 
Other photographs in the top images dataset were posted as visual imagery that suggested, in tone or content, the ``menace'' of voter fraud, or simply provided an illustration of the content of the tweet in some way. 
These illustration photos likely attempted to make use of the visual content to enhance attention, persuasion, and emotions~\cite{hsieh2011different,graber1996say,wittenberg2021minimal,counts2011taking} as well as encourage  mobilization~\cite{casas2019images}. 
We do not know if the tweets, without the illustrative images, would have been as successful. 
Such analysis is difficult to perform and control for, though past work had provided evidence that ``images evoking enthusiasm, anger, and fear should be particularly mobilizing''~\cite{casas2019images}.

We were surprised to notice several key missing or underrepresented categories amongst the popular images, most interestingly memes and manipulated photographs. 
While the presented ``evidence'' photos were clearly serving as misinformation, these photos were, to the best of our knowledge, not manipulated, but instead usually just presented out of context using the accompanied text. 
Fears of widespread use of deep fakes and other photo manipulations in the election~\cite{diakopoulos2021anticipating} did not significantly play out in this specific campaign, or at least in our dataset (more on its limitations below).
This gap, perhaps, was due to the fact that deep fakes were not needed: the people sharing the misleading content were successful in both spreading and the reception of the alleged fraud claims~\cite{CenterforanInformedPublic2021,callElectionCorrupt,FindVotesGeorgia,Berlinski,trustpoll,AmericanValuesSurvey}. 
This finding aligns with other reports (e.g.,~\cite{OutofContextPost}). 

Visual memes are another type of content that we were surprised was only sparsely represented in the set of popularly-shared images (one in our dataset of forty). 
Given the popularity of memes in other platforms~\cite{zannettou2018origins}, the role they play in political campaigns~\cite{zakem2018exploring} and in politically oriented discussion boards~\cite{ling2021dissecting}, we expected memes to be more highly represented in the data. 
We can only speculate about the reason for this absence. One such suggestion is that while memes work well on other platforms, they do not match the severity and anger that is expressed as part of the Stop the Steal campaign on Twitter.

\newt{Much of the attention in the area of visual misinformation focuses on the use of manipulated or fabricated visual content (e.g., photoshopped content, deep fakes, memes, etc.)~\cite{paris2019deepfakes,marra2018detection,marra2019gans,chesney2019deep}, or on sharing of genuine images out of their real context (e.g., sharing an image from another event, or reusing an old image)~\cite{OutofContextPost,firstdraftInformationDisorder}.
Having no manipulated photographs and only one meme in our dataset suggests that systems for detecting visual misinformation in similar contexts, i.e., political misinformation campaigns on Twitter (or other similar social media platforms), should consider investing their efforts in other attributes of the shared post such as date and account characteristics, perhaps alongside the efforts of looking for manipulated content suggested in several recent works~\cite{zhang2019detecting,xuan2019generalization,marra2019gans,diakopoulos2021anticipating,bebic2018not}. 
On the other hand, our findings do show that out-of-context images as visual misinformation, a ``cheap fake'' technique, is indeed important to address~\cite{OutofContextPost,matatov2022dataset,firstdraftInformationDisorder,matatovdejavu,moreira2018image,gupta2013faking}. 
Further, given that many of the popular images were text or quotes rendered as an image, a reasonable direction for mitigating misinformation may be using text recognition and extraction techniques to make these textual images searchable and allow detection of misinformation in tweets even when there are no suspicious keywords in the post itself.
Overall, our results can also serve as a ``compass'' for researchers and practitioners in the field of visual communication and misinformation, helping them to prioritize the most prevalent types of visual content in misinformation campaigns.}

Our analysis also focused on participation of users with different levels of popularity (in terms of number of followers). 
The results are consistent with earlier reports showing the ``multidirectional'' nature of sharing misinformation in the Stop the Steal campaign, where popular- and less-popular users participated in the spread~\cite{CenterforanInformedPublic2021}.
Moreover, our analysis showed no evidence that this co-participation changed over the lifetime of the spread of the images, with all groups of users participating in the sharing from the early instances of the image shares in our data.
We do show, however, that lower-level accounts were somewhat more involved with spreading ``evidence'' images. 
It is possible that these images simply received more attention from \newt{``spammy'' or} ``bot'' accounts, perhaps operated by disinformation actors.
\newt{However, high-follower-count accounts may also include Spam-like disinformation actors, e.g.,  ``followback networks'' where accounts follow each other to intentionally create high follower numbers, which then tend to retweet in high frequency without exercising much discretion~\cite{followback_networks}.}
Another possible explanation \newt{of the observation that high-level accounts were less involved with spreading ``evidence'' images} is that these popular users were careful about spreading more explicit misinformation, i.e. images that can get them banned from Twitter, risking loss of their followers, or simply concerned about their reputation. As a result, these users may suggest but do not state explicitly that fraud occurs -- unlike their ``foot soldiers''.
Whether they were more careful or not, our results also identified some of the most active participants amongst the very top users (those with more than 100,000 followers). We perform a novel analysis that shows whom amongst these users participated ``early and often'' in sharing the popular images. 
This analysis provided a new angle into the activity of influential repeat spreaders, known to play a role in the Stop the Steal campaign \cite{CenterforanInformedPublic2021,superspreader,gallagher2021sustained}, i.e. understanding the timing of their participation. 

Our results also expose \newt{and provide implications for} the challenging landscape for moderation of misinformation. 
Moderation can apply to content and to accounts, and may include content removal, attaching warning labels, account suspension, etc. 
Moderation is a critical tool employed by platforms to curb the spread of misinformation~\cite{gillespie2018custodians, papakyriakopoulos2020spread}, including during the U.S. 2020 elections~\cite{abilov2021voterfraud2020, Conger2021, CenterforanInformedPublic2021}.
Just like we used pHash to track appearances of the same image across tweets, moderation can make use of images to identify misinformation tweets with the same image content. 
However, as we had shown in our study, many of the popular images used in this campaign, and especially the photographic images, may not by themselves have violated content policies such as those that require identification of content manipulation. 
It was only the additional context, e.g. text associated with these images, that made them problematic. 
Even if image tracking could be successfully employed to track misinformation images, there were still a number of key challenges for moderation. 
First, as we have seen, these images spread quickly in our data, with a large majority reaching half of its shares within just a few hours, perhaps not enough (for some platforms) to make a timely decision that can help moderate their spread.
Further, while most images were shared heavily by retweeting a few original tweets, the \newt{mixed-participation pattern with shares} by users of different levels of popularity would make it hard to control the spread, \newt{e.g., by suspending selected users as moderation}. 
\newt{Nevertheless, we show that repeat and early offenders exist within the popular account group, which may offer guidelines for selective enforcement.}

Finally, we should acknowledge some limitations of this work. 
There are a number of unknown factors that potentially impact the breadth and introduce bias in the dataset we used in this work~\cite{abilov2021voterfraud2020}. 
First, the dataset was constructed by streaming from Twitter content with keywords related to ``Stop the Steal'', as defined by the authors~\cite{abilov2021voterfraud2020}. 
As a result, images that are relevant but that were not included in tweets with the correct keywords are not represented in our data.
Further, based on the analysis in the original dataset paper, the data streamed represents up to 60\% of the content posted on Twitter for these keywords~\cite{abilov2021voterfraud2020}. 
It is unknown whether there is a bias in the data that was included in the stream -- e.g. whether it was more likely to include tweets from popular users, or automatically excluded suspected spam accounts.  
Nevertheless, we do not have indication that the stream was biased in any significant way; and while it is possible that related tweets (and images) were posted without using any of the streamed keywords, we assume that a substantial fraction of the popular images were indeed shared with the popular keywords used by the Stop the Steal campaign.

\newt{Another limitation is the focus on a small number of popular images. We only looked at forty images, and patterns may be different for images that are less popular. 
However, the intent of this work was precisely to focus on the images that were shared, and seen, the most. 
Nevertheless, to evaluate the consistency and representativeness of these forty images, we also reviewed the next twenty images that were most shared by promoters of voter fraud claims. Overall, we observed similar patterns in these extra twenty images, in terms of image type, content, and the other categories that characterize them.}

\newt{We also note that the participant analysis is limited because we only have access to the number of followers of users in our dataset, and limited information about their networks beyond these counts as it is not part of the VoterFraud2020 dataset~\cite{abilov2021voterfraud2020}.
The lack of access to the set of followers for each participating account prevents us from calculating the overlap between the followers of different accounts, which in turn greatly inflates the images followers' estimated exposure values. 
Thus, our potential follower exposure analysis presents an upper bound for the actual half-life of the images' exposure.
}

Of course, our results are limited to a single social media platform, and single campaign -- though clearly an influential and consequential one~\cite{FindVotesGeorgia,callElectionCorrupt, benkler2020mail, CenterforanInformedPublic2021}.   
Still, we do not expect that our precise findings will generalize to other platforms and settings. 
However, the methods we used in this work can be applied in other contexts, and the insights provided about the use of images in this specific campaign contribute to inform our understanding of the Stop the Steal campaign with its consequences, which are still playing out at the time of publication, and will continue to impact the state of democracy in the U.S. and beyond for a long time to come. %

\begin{acks}
This material is based upon work partially supported by the National Science Foundation under grants SaTC-2120651 and IIS-1840751. We thank the Jacobs Technion-Cornell Institute for supporting this project.
\end{acks}

\appendix
\section{Appendix A: Image Dataset Details}
\label{appendix:images}
\begin{table}[H]
\centering
\tiny
\begin{tabular}{|c|c|c|c|c|c|}
\hline
\textbf{\begin{tabular}[c]{@{}c@{}}Image Popularity\\ Rank\end{tabular}} & \textbf{\begin{tabular}[c]{@{}c@{}}Retweets by\\ Promoters\end{tabular}} & \textbf{\begin{tabular}[c]{@{}c@{}}Retweets by\\ Detractors\end{tabular}} & \textbf{\begin{tabular}[c]{@{}c@{}}Total Retweets \\ in the Dataset\end{tabular}} & \textbf{\begin{tabular}[c]{@{}c@{}}Total Tweets\\ in the Dataset\end{tabular}} & \textbf{\begin{tabular}[c]{@{}c@{}}Total Retweets\\ According to Metadata\end{tabular}} \\ \hline
1 & 10424 & 20 & 12995 & 11 & 20104 \\ \hline
2 & 10250 & 140 & 12370 & 34 & 28833 \\ \hline
3 & 7076 & 16 & 8847 & 9 & 13769 \\ \hline
4 & 5698 & 42 & 7524 & 9 & 18979 \\ \hline
5 & 5619 & 23 & 6513 & 31 & 15896 \\ \hline
6 & 5521 & 21 & 6385 & 19 & 15597 \\ \hline
7 & 5518 & 21 & 6378 & 12 & 15594 \\ \hline
8 & 4902 & 54 & 6335 & 9 & 19403 \\ \hline
9 & 4090 & 43 & 5053 & 14 & 13414 \\ \hline
10 & 3924 & 12 & 4837 & 45 & 32050 \\ \hline
11 & 3288 & 9 & 4015 & 2 & 5547 \\ \hline
12 & 2619 & 5 & 3203 & 100 & 8699 \\ \hline
13 & 2012 & 3 & 2345 & 36 & 6349 \\ \hline
14 & 1992 & 0 & 2451 & 6 & 3020 \\ \hline
15 & 1980 & 3 & 2372 & 23 & 3748 \\ \hline
16 & 1965 & 1 & 2465 & 1 & 3968 \\ \hline
17 & 1954 & 0 & 2236 & 2 & 2900 \\ \hline
18 & 1898 & 7 & 2373 & 2 & 4394 \\ \hline
19 & 1805 & 3 & 2148 & 1 & 4033 \\ \hline
20 & 1765 & 4 & 2164 & 1 & 2501 \\ \hline
21 & 1724 & 4 & 2061 & 2 & 4738 \\ \hline
22 & 1628 & 2 & 2053 & 1 & 2202 \\ \hline
23 & 1571 & 4 & 2085 & 11 & 2261 \\ \hline
24 & 1512 & 3 & 1792 & 4 & 6110 \\ \hline
25 & 1507 & 0 & 1607 & 5 & 1690 \\ \hline
26 & 1494 & 3 & 1772 & 1 & 6068 \\ \hline
27 & 1466 & 7 & 1828 & 1 & 3262 \\ \hline
28 & 1416 & 0 & 1507 & 6 & 1588 \\ \hline
29 & 1405 & 1 & 1516 & 6 & 1639 \\ \hline
30 & 1374 & 6 & 1589 & 1 & 1764 \\ \hline
31 & 1374 & 2 & 1840 & 1 & 2046 \\ \hline
32 & 1374 & 6 & 1589 & 1 & 1764 \\ \hline
33 & 1366 & 1 & 1746 & 3 & 2109 \\ \hline
34 & 1348 & 5 & 1580 & 1 & 1661 \\ \hline
35 & 1339 & 2 & 1690 & 1 & 1799 \\ \hline
36 & 1332 & 0 & 1436 & 1 & 3634 \\ \hline
37 & 1332 & 0 & 1436 & 1 & 3634 \\ \hline
38 & 1148 & 4 & 1523 & 4 & 1734 \\ \hline
39 & 1099 & 0 & 1399 & 4 & 1557 \\ \hline
40 & 1035 & 5 & 1231 & 1 & 4975 \\ \hline
\end{tabular}
\caption{The forty images used in our data, i.e. the images most shared by promoters of voter fraud claims, and their data.}
\label{table:top_40_images}
\end{table}

\bibliographystyle{ACM-Reference-Format}
\bibliography{bib}

\received{January 2022}
\received[revised]{April 2022}
\received[accepted]{August 2022}

\end{document}